# AHB Compatible DDR SDRAM Controller IP Core for ARM BASED SOC

[1]Dr.R.Shashikumar, [2]C.N. Vijay Kumar, [3]M.Nagendrakumar, [4]C.S.Hemanthkumar
Professor      Asst.prof         Asst.Prof         Sr.Lecturer
ECE dept, SJCIT, Chikkaballapur, Karnataka, India     JVIT, Bidadi
,

*Abstract*— DDR SDRAM is similar in function to the regular SDRAM but doubles the bandwidth of the memory by transferring data on both edges of the clock cycles. DDR SDRAM most commonly used in various embedded application like networking, image/video processing, Laptops etc. Now a day's many applications needs more and more cheap and fast memory. Especially in the field of signal processing, requires significant amount of memory. The most used type of dynamic memory for that purpose is DDR SDRAM. For FPGA design the IC manufacturers are providing commercial memory controller IP cores working only on their products. Main disadvantage is the lack of memory access optimization for random memory access patterns. The 'data path' part of those controllers can be used free of charge. This work propose an architecture of a DDR SDRAM controller, which takes advantage of those available and well tested data paths and can be used for any FPGA device or ASIC design.[5]. In most of the SOC design, DDR SDRAM is commonly used. ARM processor is widely used in SOC's; so that we focused to implement AHB compatible DDR SDRAM controller suitable for ARM based SOC design.

## 1 INTRODUCTION

The DDR SDRAM is a high-speed CMOS, dynamic random-access memory. It is internally configured as a quad bank DRAM. The DDR SDRAM uses double data rate architecture to achieve high-speed operation. The double data rate architecture is essentially 2n prefetch architecture with an interface designed to transfer two data words per clock cycle at the I/O pins. A single read or write access for the DDR SDRAM effectively consists of a single 2*n*-bit wide, one-clock-cycle data transfer at the internal DRAM core and two corresponding *n*-bit wide, one-half clock-cycle data transfers at the I/O pins. A bidirectional data strobe (DQS) is transmitted externally, along with data, for use in data capture at the receiver. DQS is a strobe transmitted by the DDR SDRAM during READs and by the memory controller during WRITEs. DQS is edge-aligned with data for READs and center-aligned with data for WRITEs.

Read and write accesses to the DDR SDRAM are burst oriented; accesses start at a selected location and continue for a programmed number of locations in a programmed sequence. Accesses begin with the registration of an ACTIVE command, which is then followed by a READ or WRITE command. The address bits registered coincident with the ACTIVE command are used to select the bank and row to be accessed. The address bits registered coincident with the READ or WRITE command are used to select the bank and the starting column location for the burst access.

The DDR SDRAM provides for programmable READ or WRITE burst lengths of 2, 4, or 8 locations. An auto precharge function may be enabled to provide a self timed row precharge that is initiated at the end of the burst access This model has implemented in RTL by Verilog. The focus of this work is to implement behavioral model of DDR SDRAM and also implemented on the Xilinx Spartan series FPGA.The Top level model is as shown in Fig.1. The core contains mainly two parts, AHB Slave and DDR SDRAM controller.

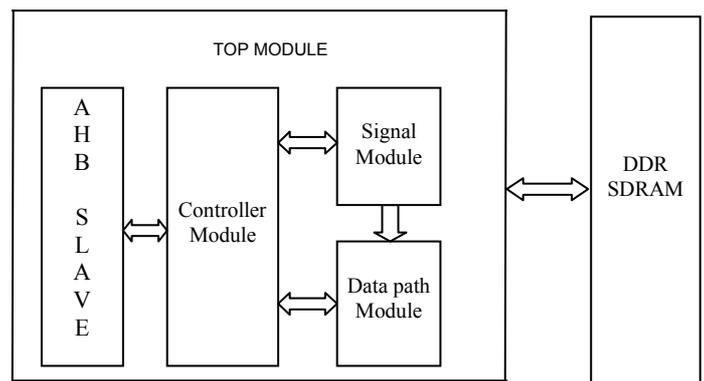

Figure .1 Top Module

## 2. FUNCTIONAL BLOCK DIAGRAM

The functional block diagram of the DDR controller is shown in Fig.2. It consists of four modules, AHB Salve, the main control module, the signal generation module and the data path module. The AHB slave normally connected to the AHB bus Arbiter in the design. AHB master sends data to the AHB slave based on that protocol. Burst mode read and writes and split transaction read and writes transactions supported. AHB slave complies with the processor interface protocol of ARM processor. The main control module has two state machines and a refresh counter, which generates proper istate and cstate outputs according to the system interface control signals. The signal generation module generates the address and command signals required for DDR based on istate and cstate. The data path module performs the data latching and dispatching of the data between the Processor and DDR.

The DDR SDRAM provides for programmable READ or WRITE burst lengths of 2, 4, or 8 locations. An AUTO PRECHARGE function may be enabled to provide a self-timed row precharge that is initiated at the end of the burst access. As with standard SDR SDRAMs, the pipelined, multibank architecture of DDR SDRAMs allows for concurrent operation, thereby providing high effective bandwidth by hiding row precharge and activation time. An auto refresh mode is provided, along with a power-saving power-down mode.

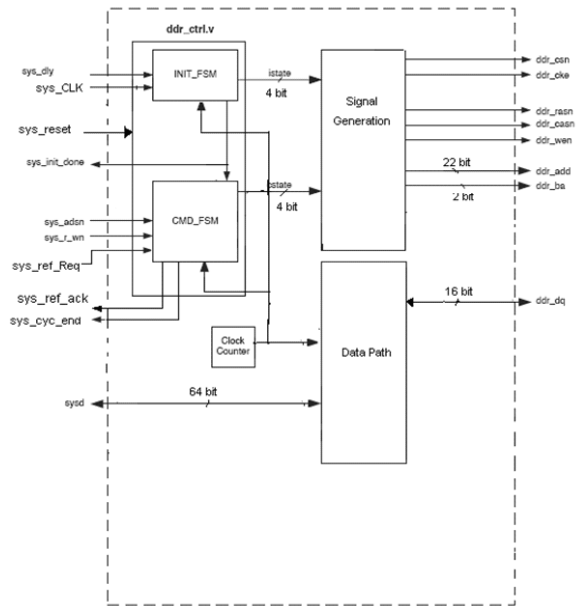

Figure .2 Functional Block Diagram

3. FUNCTIONAL DESCRIPTION

3.1. Initialization

DDR SDRAMs must be powered up and initialized in a predefined manner. Operational procedures other than those specified may result in undefined operation.
DDR SDRAM requires a 200μs delay prior to applying an executable command [7]. Once the 200μs delay has been satisfied, a DESELECT or NOP command should be applied. Following the NOP command, a PRECHARGE ALL command should be applied. Next a LOAD MODE REGISTER command should be issued. A PRECHARGE ALL command should then be applied, placing the device in the all banks idle state. Once in the idle state, two AUTO REFRESH cycles must be performed (tRFC must be satisfied.) Additionally, a LOAD MODE REGISTER command for the mode register is issued [16].

3.2 Register Definition

The mode register is used to define the specific mode of operation of the DDR SDRAM. This definition includes the selection of a burst length, a burst type, a CAS latency and an operating mode. The mode register is programmed via the MODE REGISTER SET command (with BA0 = 0 and BA1 = 0) and will retain the stored information until it is programmed again or the device loses power (except for bit A8, which is self-clearing). Reprogramming the mode register will not alter the contents of the memory, provided it is performed correctly. The mode register must be loaded (reloaded) when all banks are idle and no bursts are in progress, and the controller must wait the specified time before initiating the subsequent operation. Violating either of these requirements will result in unspecified operation. Mode register bits A0-A2 specify the burst length, A3 specifies the type of burst (sequential or interleaved), A4-A6 specifies the CAS latency, and A7-A12 specifies the operating mode. Fig.3 shows the description of mode register.

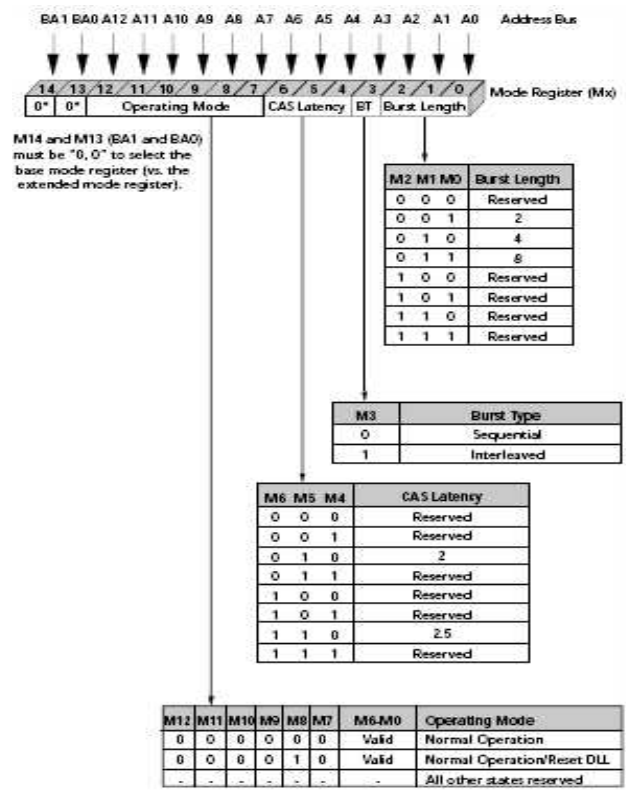

Figure.3 Mode register definition

3.3 DDR Commands

Table 1 presents the commands issued by the controller. These commands are passed to the memory using the following control signals[7]:
• Row Address Select (RAS)
• Column Address Select (CAS)
• Write Enable (WE)
• Clock Enable (CKE) (always held High after device configuration)
• Chip Select (CS) (always held Low during device operation)

| Signal No. | Function | RAS | CAS | WE |
|---|---|---|---|---|
| 1 | Load Mode Register | L | L | L |
| 2 | Auto Refresh | L | L | H |
| 3 | Precharge(1) | L | H | L |
| 4 | Select Bank Activate Row | L | H | H |
| 5 | Write Command | H | L | L |
| 6 | Read Command | H | L | H |

| Signal No. | Function | RAS | CAS | WE |
|---|---|---|---|---|
| 7 | No Operation (NOP) | H | H | H |

Table 1 DDR SDRAM Commands

3.3.1 Command Functions

The fallowing commands are used in DDR SDRAM controller core.

- Mode Register :

The Mode register is used to define the specific mode of DDR SDRAM operation, including the selection of burst length, burst type, CAS latency, and operating mode.

- Auto Refresh :

The REFRESH command instructs the controller to perform an AUTO REFRESH command to the SDRAM. The controller will acknowledge the REFRESH command with ACK. DDR SDRAM is somewhat similar to regular SDRAM. Both will break the RAM into smaller chunks for simultaneous, synchronized request-and-reply access. In addition, both types of memory can be packaged in DIMM modules. However, DDR SDRAM will perform the alternating request-and-reply rhythm on both the rise and fall of the clock cycle. This method effectively doubles the bandwidth available and increases the speed the system can access data in memory

- Precharge:

The PRECHARGE command is used to deactivate the open row in a particular bank. The bank is available for subsequent row activation for a specified time (tRP) after the PRECHARGE command is issued. Input A10 determines whether one or all banks are precharged.

- ACTIVE Command:

The ACTIVE command activates a row in a bank, allowing any READ or WRITE commands to be issued to a bank in the memory array. After a row has been opened, READ or WRITE commands can be issued to that row, subject to the tRCD specification. When the controller detects an incoming address that refers to a row in a bank other than the currently opened row, the controller issues an address conflict signal. A PRECHARGE command is also issued by the controller to deactivate the open row. The controller also issues another ACTIVE command to the new row.

- READ Command:

The READ command is used to initiate a burst read access to an active row. The value on BA0 and BA1 selects the bank address. The address inputs provided on A0 – Ai select the starting column location. After the read burst is over, the row is still available for subsequent access until it is precharged.

- WRITE Command:

The WRITE command is used to initiate a burst access to an active row. The value on BA0 and BA1 selects the bank address, while the value on address inputs A0 – Ai selects the starting column location in the active row. The value of Write Latency is equal to one clock cycle.

4 .DDR SDRAM Controller Block

The controller block mainly consists of two FSMs
- Initial FSM
- Command FSM

It generates the required commands for initializing the DDR SDRAM. The block diagram for the DDR SDRAM controller is as shown in Fig.4.

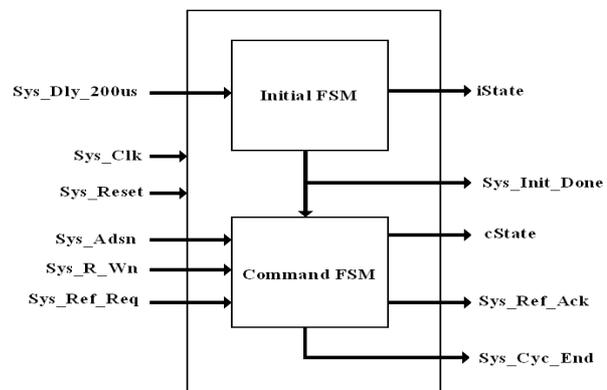

Figure 4 Block Diagram of Controller Block

4.1 DDR SDRAM Initial FSM

Before normal memory accesses can be performed, DDR needs to be initialized by a sequence of commands. The INIT_FSM state machine handles this initialization. Fig.5 shows the state diagram of the INIT_FSM state machine. During reset, the INIT_FSM is forced to the i_IDLE state. After reset, the sys_dly_200US signal will be sampled to determine if the 200μs power/clock stabilization delay is completed. After the power/clock stabilization is complete, the DDR initialization sequence will begin and the INIT_FSM will switch from i_IDLE to i_NOP state and in the next clock to I_PRE.
The initialization starts with the PRECHARGE ALL command. Next a LOAD MODE REGISTER command will be applied for the extended mode register to enable the DLL inside DDR, followed by another LOAD MODE REGISTER command to the mode register to reset the DLL. Then a PRECHAGE command will be applied to make all banks in the device to idle state. Then two, AUTO REFRESH commands, and then the LOAD MODE REGISTER command to configure DDR to a specific mode of operation. After issuing the LOAD MODE REGISTER command and the tMRD timing delay is satisfied, INIT_FSM goes to i_ready state and remains there for the normal memory access cycles unless reset is asserted. Also, signal sys_init_done is set to high to indicate the DDR initialization is completed. The i_PRE, i_AR1, i_AR2, i_EMRS and i_MRS states are used for issuing DDR commands. The LOAD MODE REGISTER command configures the DDR by loading data into the mode register through the address bus. The data present on the address bus (ddr_add) during the LOAD MODE REGISTER command is loaded to the mode register. The mode register contents specify the burst length, burst type, CAS latency, etc. A PRECHARGE/AUTO PRECHARGE command moves all banks to idle state. As long as all banks of the DDR are in idle state, mode register can be reloaded with different value thereby changing the mode of operation. However, in most applications the mode register value will not be changed after initialization. This design assumes the mode register stays the same after initialization.

4.1.1 Initial FSM State Diagram:

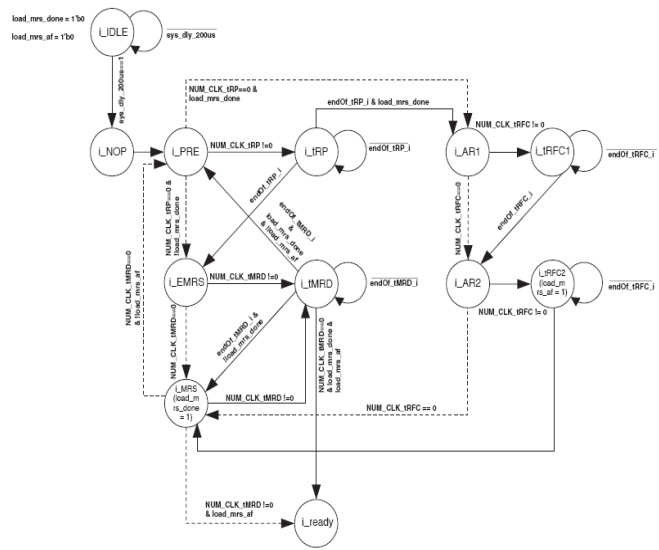

Figure .5 Initial FSM State Diagram

As mentioned above, certain timing delays (tRP, tRFC, tMRD) need to be satisfied before another non-NOP command can be issued. These SDRAM delays vary from speed grade to speed grade and sometimes from vendor to vendor. To accommodate this without sacrificing performance, the designer needs to modify the HDL code for the specific delays and clock period (tCK). According to these timing values, the number of clocks the state machine will stay at i_tRP, i_tRFC1, i_tRFC2, i_tMRD states will be determined after the code is synthesized. In cases where tCK is larger than the timing delay, the state machine doesn't need to switch to the timing delay states and can go directly to the command states[15]. The dashed lines in Fig 5 show the possible state switching paths.

4.1.2 Different states of Initial FSM

*a) Idle:*

When reset is applied the initial fsm is forced to IDLE state irrespective of which state it is actually in when system is in idle it remains idle without performing any operations.

*b) No Operation:*

The NO OPERATION (NOP) command is used to instruct the selected DDR SDRAM to perform a NOP (CS# is LOW with RAS#, CAS#, and WE# are HIGH). This prevents unwanted commands from being registered during idle or wait states. Operations already in progress are not affected.

*c) Precharge:*

The PRECHARGE command is used to deactivate the open row in a particular bank or the open row in all banks as shown in Fig.6. The value on the BA0, BA1 inputs selects the bank, and the A10 input selects whether a single bank is precharged or whether all banks are precharged.

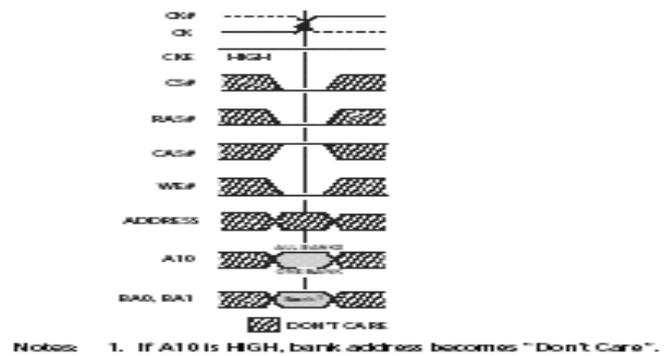

Figure .6 Precharge Commands.

*d) Auto Refresh:*

AUTO REFRESH is used during normal operation of the DDR SDRAM and is analogous to CAS#-before-RAS# (CBR) refresh in DRAMs. This command is nonpersistent, so it must be issued each time a refresh is required. All banks must be idle before an AUTO REFRESH command is issued.

*e) Load Mode register(LMR):*

The mode registers are loaded via inputs A0–A*n*. The LOAD MODE REGISTER command can only be issued when all banks are idle, and a subsequent executable command cannot be issued until tMRD is met.

*f) Read/Write Cycle:*

The Fig.5 shows the state diagram of CMD_FSM which handles the read, write and refresh of the SDRAM.

The CMD_FSM state machine is initialized to c_idle during reset. After reset, CMD_FSM stays in c_idle as long as sys_INIT_DONE is low which indicates the SDRAM initialization sequence is not yet completed. Once the initialization is done, sys_ADSn and sys_REF_REQ will be sampled at the rising edge of every clock cycle. A logic high sampled on sys_REF_REQ will start a SDRAM refresh cycle. This is described in the following section. If logic low is sampled on both sys_REF_REQ and sys_ADSn, a system read cycle or system write cycle will begin. These system cycles are made up of a sequence of SDRAM commands. Initialization:

Prior to normal operation, DDR SDRAMs must be powered up and initialized in a predefined manner. Operational procedures, other than those specified, may result in undefined operation.

### 4.2 DDR SDRAM COMMAND FSM

The fig.7 shows the state diagram of CMD_FSM, which handles read, write and refresh of the DDR. The CMD_FSM state machine is initialized to c_idle during reset. After reset, CMD_FSM stays in c_idle as long as sys_init_done is low which indicates th e DDR initialization sequence is not yet completed. From this state, a READA/WRITEA/REFRESH cycle starts depending upon sys_adsn/ rd_wr_req_during_ ref_req signals as shown in the state diagram. All rows are in the "closed" status after the DDR initialization. The rows need to be "opened" before they can be accessed. However, only one row in the same bank can be opened at a time. Since there are four banks, there can be at most four rows opened at the same time. If a row in one bank is currently opened, it needs to be closed before another row in the same bank can be opened. ACTIVE command is used to open the rows and PRECHARGE is used to close the rows. When issuing the commands for opening or closing the rows, both row address and bank address need to be provided.

In this design, the ACTIVE command will be issued for each read or write access to open the row. After a tRCD delay is satisfied, READA or WRITEA commands will be issued with a high ddr_add[10] to enable the AUTO REFRESH for closing the row after access. Therefore, the clocks required for read/write cycle are fixed and the access can be random over the full address range. Read or write is determined by the sys_r_wn status sampled at the rising edge of the clock before the tRCD delay is satisfied. If logic high is sampled, the state machine switches to c_READA. If a logic low is sampled, the state machine switches to c_WRITEA.

For read cycles, the state machine switches from, c_READA to c_cl for CAS latency, then switches to crate for transferring data from DDR to processor. The burst length determines the number of clocks the state machine stays in c_rdata state. After the data is transferred, it switches back to c_idle.

For write cycles, the state machine switches from c_WRITEA to c_wdata for transferring data from bus master to DDR, then switches to c_tDAL. After the clock rising edge of the last data in the burst sequence, no commands other than NOP can be issued to DDR before tDAL is satisfied.

4.2.1 Different states of Command FSM

*a) Refresh Cycle:*

DDR memory needs a periodic refresh to hold the data. This periodic refresh is done using AUTO REFRESH command. All banks must be idle before an AUTO REFRESH command is issued. In this design all banks will be in idle state, as every read/write operation uses auto pre charge.

*b) Active:*

The ACTIVE command is used to open (or activate) a row in a particular bank for a subsequent access, like a read or a write, as shown in Fig.8. The value on the BA0, BA1 inputs selects the bank, and the address provided on inputs A0–A*n* selects the row.

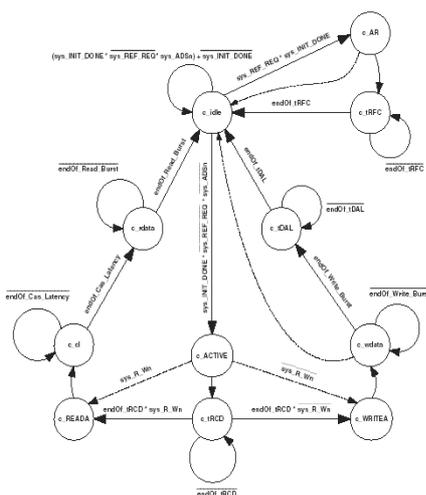

Figure .7 Command FSM State Diagram

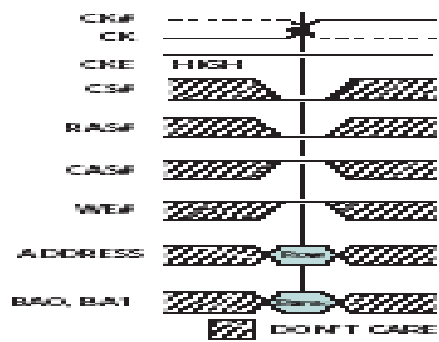

Figure.8 Activating a Specific Row in a Specific Bank.

c) Read:

The READ command is used to initiate a burst read access to an active row, as shown in Fig.8. The value on the BA0, BA1 inputs selects the bank, and the address provided on inputs A0–A$i$ (where A$i$ is the most significant column address bit for a given density and configuration) selects the starting column location.

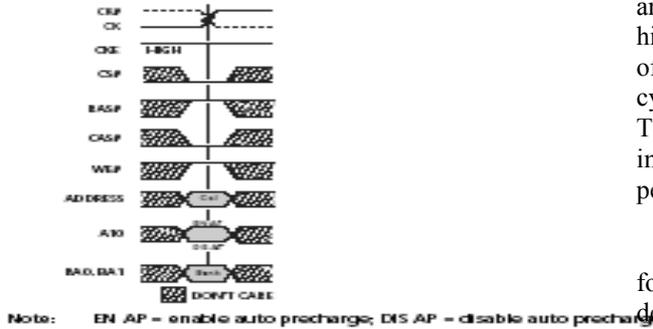

Figure .9 Read Command

d) Write:

The WRITE command is used to initiate a burst write access to an active row as shown in Fig.10. The value on the BA0, BA1 inputs selects the bank, and the address provided on inputs A0–A$i$ (where A$i$ is the most significant column address bit for a given density and configuration) selects the starting column location.

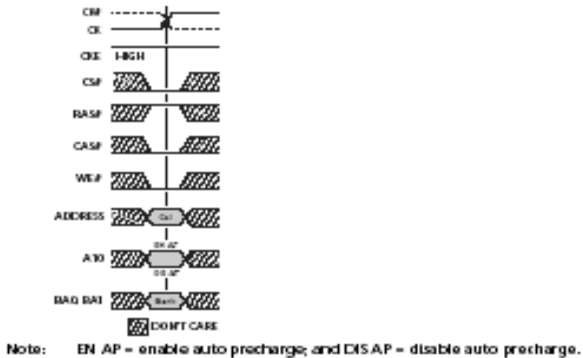

Figure .10 Write Command

Similar to the FP and EDO DRAM, row address and column address are required to pinpoint the memory cell location of the SDRAM access. Since SDRAM is composed of four banks, bank address needs to be provided as well.

The SDRAM can be considered as a four by N array of rows. All rows are in the "closed" status after the SDRAM initialization. The rows need to be "opened" before they can be accessed. However, only one row in the same bank can be opened at a time. Since there are four banks, there can be at most four rows opened at the same time. If a row in one bank is currently opened, it must be closed before another row in the same bank can be opened. ACTIVE command is used to open the rows and PRECHARGE (or the AUTO PRECHARGE hidden in the WRITE and READ commands, as used in this design) is used to close the rows. When issuing the commands for opening or closing the rows, both row address and bank address need to be provided.

For sequential access applications and those with page memory management, the proper address assignments and the use of the SDRAM pipeline feature deliver the highest performance SDRAM controller. However, this type of controller design is highly associated with the bus master cycle specification and will not fit the general applications. Therefore, this SDRAM controller design does not implement these custom features to achieve the highest performance through these techniques.

In this design, the ACTIVE command will be issued for each read or write access to open the row. After a tRCD delay is satisfied, READ or WRITE commands will be issued with a high sdr_A[10] to enable the AUTO REFRESH for closing the row after access. So, the clocks required for read/write cycle are fixed and the access can be random over the full address range. Read or write is determined by the sys_R_Wn status sampled at the rising edge of the clock before tRCD delay is satisfied. If a logic high is sampled, the state machine switches to c_READA. If a logic low is sampled, the state machine switches to c_WRITEA.For read cycles, the state machine switches from c_READA to c_cl for CAS latency, then switches to c_rdata for transferring data from SDRAM to bus master. The number of clocks the state machine stays in c_rdata state is determined by the burst length. After the data is transferred, it switches back to c_idle.For write cycles, the state machine switches from c_WRITEA to c_wdata for transferring data from bus master to SDRAM, then switches to c_tDAL. Similar to read, the number of clocks the state machine stays in c_wdata state is determined by the burst length. The time delay tDAL is the sum of WRITE recovery time tWR and the AUTO PRECHARGE timing delay tRP. After the clock rising edge of the last data in the burst sequence, no commands other than NOP can be issued to SDRAM before tDAL is satisfied.As mentioned in the INIT_FSM section above, the dash lines indicates possible state switching paths when tCK period is larger than timing delay spec.

e) Refresh cycle:

Similar to the other DRAMs, memory refresh is required. A SDRAM refresh request is generated by activating sdr_REF_REQ signal of the controller. The sdr_REF_ACK signal will acknowledge the recognition of sdr_REF_REQ and will be active throughout the whole refresh cycle. The sdr_REF_REQ signal must be maintained until the sdr_REF_ACK goes active in order to be recognized as a refresh cycle. Note that no system read/write access cycles are allowed when sdr_REF_ACK is active. All system interface cycles will be ignored during this period. The

sdr_REF_REQ signal assertion needs to be removed upon receipt of sdr_REF_ACK acknowledge, otherwise another refresh cycle will again be performed.

Upon receipt of sdr_REF_REQ assertion, the state machine CMD_FSM enters the c_AR state to issue an AUTO REFRESH command to the SDRAM. After tRFC time delay is satisfied, CMD_FSM returns to c_idle.

## 5. DATA PATH

The fig.11 shows the data path module with inputs and outputs as shown in the figure

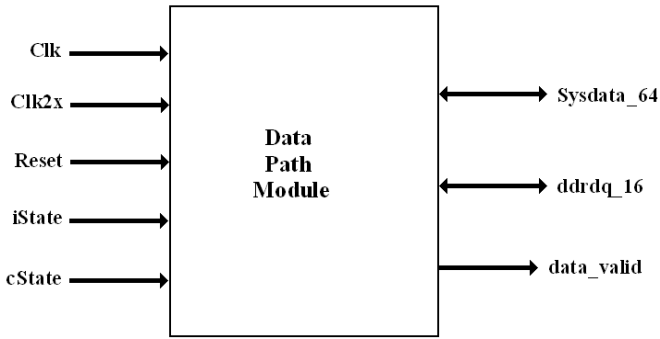

Figure .11.Data Path Module.

The data flow design between the SDRAM and the system interface. The module in this reference design interfaces between the SDRAM with 16-bit bidirectional data bus and the bus master with 64-bit bidirectional data bus. The user should be able to modify this module to customize to fit his/her system bus requirements. The data path module performs the data latching and dispatching of the data between the processor and DDR.

## 6. TIMING DIAGRAMS

The fig.12 and Fig.13 are the read cycle and write cycle timing diagrams of the reference design with the two CAS latency cycles and the burst length of four. In the example shown in the figures, the read cycle takes 10 clocks and the write cycle takes 9 clocks. The state variable c_State of CMD_FSM is also shown in these figures. Note that the ACTIVE, READ, WRITE commands are asserted one clock after the c_ACTIVE, c_READA, c_WRITEA states respectively. The values of the region filed with slashes in the system interface input signals of these figures are "don't care." For example, signal sys_R_Wn needs to be valid only at the clock before CMD_FSM switches to the c_READA or c_WRITEA states. Depending on the values of tRCD and tCK, this means the signal sys_R_Wn needs to be valid at state c_ACTIVE or the last clock of state c_tRCD.

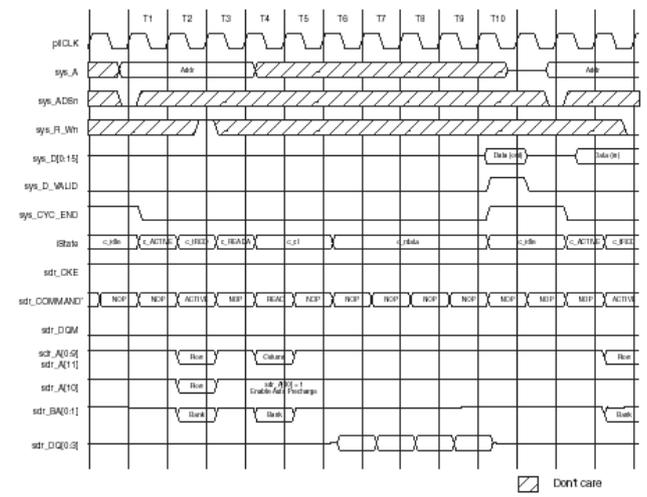

Figure .12 Read Cycle Timing Diagram

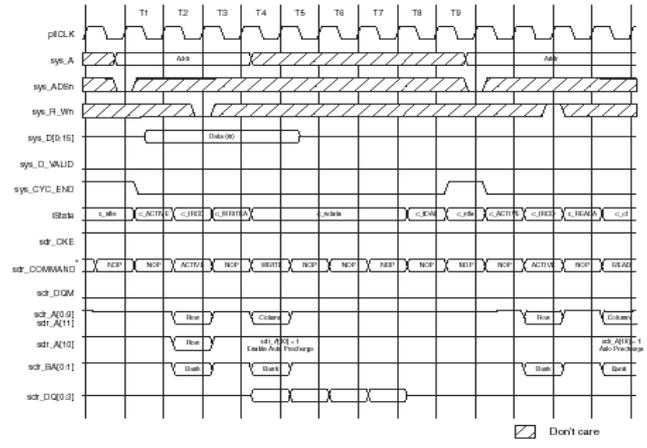

Figure.13 Write Cycle Timing Diagram.

The signal generation module generates the address and command signals required for DDR based on istate and cstate

## 6. IMPLEMENTATION AND RESULT

DUT mainly contains two parts, AHB slave and DDR SDRAM controller. The core is developed by using verilog [2]. In this work we have designed a High speed DDR SDRAM Controller with 64-bit data transfer which synchronizes the transfer of data between DDR RAM and External AHB compatible devices. This can be used in ARM based SOC design. The advantages of this controller compared to SDR SDRAM is that it synchronizes the data transfer, and the data transfer is twice as fast as previous, the production cost is also very low. This core is verified by using testbench and several testcases, which cover most of the functionality of the core. The simulations of specified functions were conducted by the Modelsim tool [3].The fig.14, Fig.15 and Fig 16 shows the waveforms initial, command and data path of the core. Fig.17 and 18 shows the RTL schematic of top module and controller respectively.

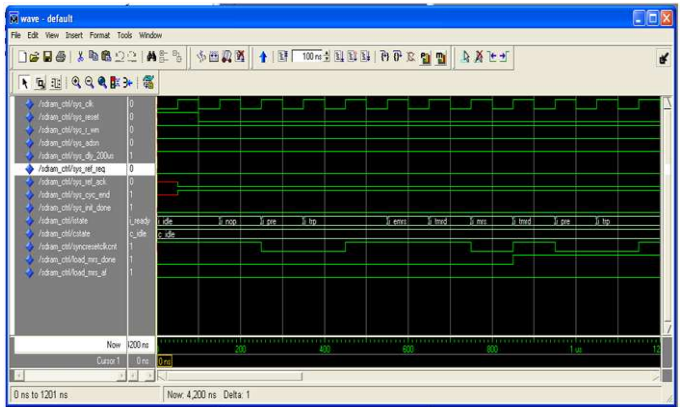

Figure.14 Waveforms of initial FSM

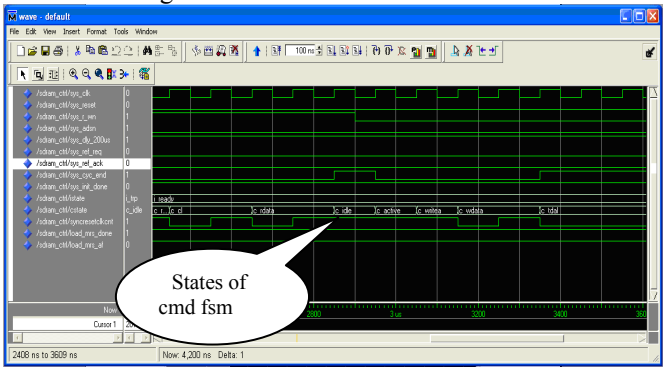

Figure.15 Waveforms of Command FSM

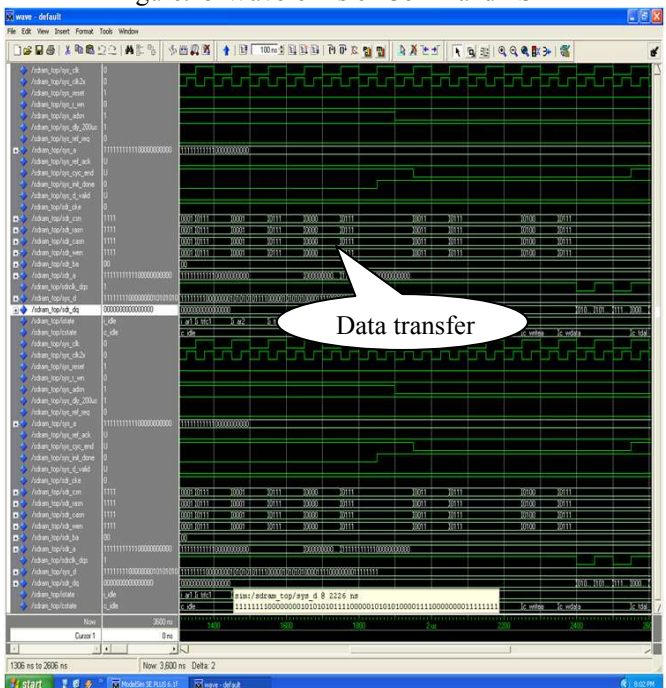

Figure .16 Waveforms Of Data path

For verification of core, two BFM's are used, AHB BFM at the input side and DDR memory model BFM at the output side. Core has been tested for main functionality of DDR SDRAM controller. AHB slave tested for both split and burst transfers. Our implementation of the design was analyzed by using ISE FPGA tool from Xilinx [1]. Designs are mapped on to Spartan FPGA. The whole system has been placed and routed in to the XC3S500e-5-PQ208 FPGA chip. The numbers of slices utilized by the design are 1340 out of 4656, 28% of the utilization.

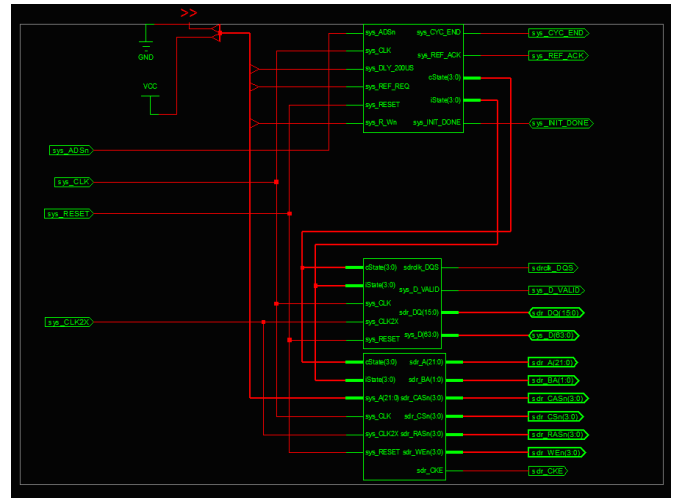

Figure .17.RTL Schematic of Top Module

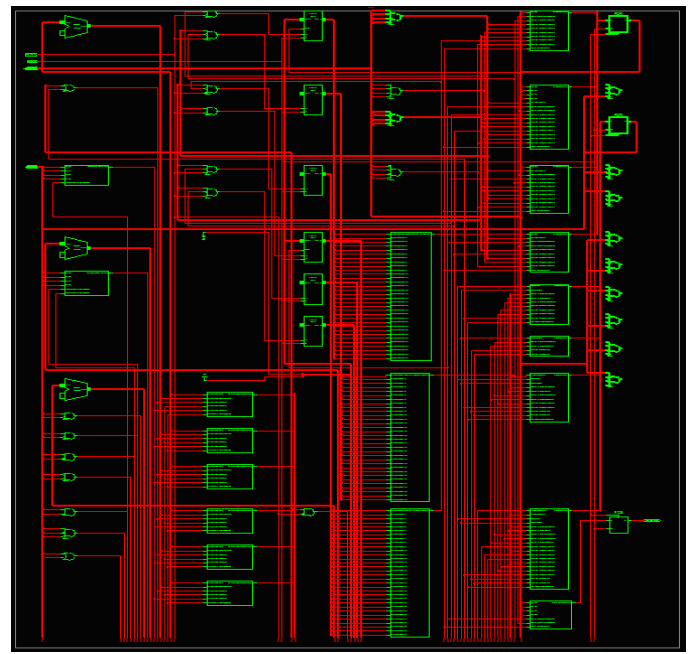

Figure .18.RTL Schematic of Controller

HDL Synthesis report of the core as shown below
**Macro Statistics**
# Adders/Subtractors         : 7
# 3-bit adder                : 2
#32-bit subtractor           : 6
# Counters                   : 7
# 32-bit down counter        : 7

```
# Registers                : 18
# 1-bit register           : 8
# 16-bit register          : 2
#3-bit register            : 1
# 32-bit register          : 4
#4-bit register            : 2
# Latches                  : 2
# 64-bit latch             : 1
```
---------------------------
**Device utilization summary:**
---------------------------
**Selected Device : 3s400tq144-5**

```
Number of Slices         :  482 out of  3584    14%
Number of Slice Flip Flops :  468 out of  7168    6%
Number of 4 input LUTs   :  8944 out of 7168    12%
Number of IOs            :  165
Number of bonded IOBs    :  165 out of   97    629% (*)
IOB Flip Flops           :  71
Number of GCLKs          :  3 out of    8      37%
```
**Timing Summary:**
Speed Grade:     -5

Minimum period: 9.802ns (Maximum Frequency: 102.021MHz)

Minimum input arrival time before clock: 5.954ns Maximum output required time after clock: 10.704ns

Maximum combinational path delay: 8.451ns

DDR2 SDRAM is the second generation of DDR SDRAM. DDR2 SDRAM improves on DDR SDRAM by using differential signaling and lower voltages to support significant performance advantages over DDR SDRAM. DDR SDRAM standards are still being developed and improved [20].

## AUTHORS PROFILE

[1] Dr. R. Shashikumar is presently working as a Professor 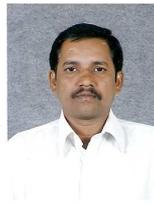 in E & C dept, SJCIT, Chikballapur, Karnataka, India. He is having 10 years of teaching and 6 years of Industry experience. His areas of interest includes ASIC, FPGA, Network Security.

[2] Prof.C.N.Vijayakumar [M.E, MISTE, MIE, MIETE] is presently working 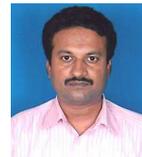 as a HOD and  Assistant Professor in the department of Telecommunication engg , SJCIT, Chikballapur, Karnataka, India. He is having 15 years of teaching experience. His areas of interest are Power Electronics, Low Power VLSI, ASIC and Control System.

[3]Mr. M.N.NagendraKumar [M.E, MISTE] is working as a Asst.Professor in the Dept of E & C, SJCIT, Chikballapur, Karnataka, India. He is having 13 years of teaching experience. His areas of interest are Power electronics, Control System and VLSI.

[4]Mr. C.S.Hemanthkumar [M.Tech, MISTE] is working as a Sr.Lecturer in the Dept of E&C, JVIT, Bangalore, India. His areas of interest are VLSI, Embedded, Signal Processing.